# A Comprehensive Study of the Physical and Geometrical Characteristics of the Close Visual Binary System HIP 45571

Motasem J. Alslaihat[1,4(✉)], Hatem S. Widyan[1], Mashhoor A. Al-Wardat[1,2,3,4(✉)], Awni M. Kasawneh[1,2,3,4], Diala M. Taneenah[1,4], and Abdullah M. Hussein[1,4]

[1] Department of Physics, Al Al-Bayt University, Mafraq 25113, Jordan
Alabbade47@gmail.com, malwardat@sharjah.ac.ae
[2] Department of Applied Physics and Astronomy, University of Sharjah, Sharjah 27272, UAE
[3] Space Sciences and Technology, Sharjah Academy for Astronomy, University of Sharjah, Sharjah 27272, UAE
[4] Arab Union for Astronomy and Space Sciences, Amman 11941, Jordan

**Abstract.** In this paper, we estimated the physical and geometrical characteristics of the visually close binary stellar system Hip 45571, using "Al-Wardat's method for analyzing binary and multiple stellar systems". We estimated the physical properties of the components of the system for the four measured parallax given by Gaia and Hipparcos, which gives a dynamical mass sum ranges between 2.43 and 2.52 solar mass using the new orbital parameters following Tokovinin's dynamical method.

The method used is a spectrophotometrical computational technique that employs Kurucz plane-parallel line-blanketed model atmospheres for single stars. These model atmospheres are used to construct the synthetic spectral energy distributions (SED) of each component and for the entire system. To ensure the method's accuracy, we apply the fit between synthetic and observational photometry under different filters, including the recently published Gaia DR3 measurements. The positions of the components on the H-R diagram and the evolutionary tracks were used to estimate their masses and ages.

We found that the system consists of 2.24 Gyr two F2.5 IV and F3.5IV subgiant components with $T_{\text{eff}}^A = 6800$ K, $T_{\text{eff}}^B = 6700$ K, $\log g_A = 4.19$ m/s$^2$ $\log g_B = 4.33$ m/s$^2$, $R_A = 1.77 R_\odot$, $R_B = 1.34 R_\odot$, $L_A = 6.01 L_\odot$, $L_B = 3.25 L_\odot$ and, $Z_* = 0.011$.

Depending on the masses estimated by Al-Wardat's method, a new parallax value of $28.72 \pm 0.30$ mas was obtained. Which lies between the values given by DR2 and DR3. This research underscores the importance of precision and reliability in employing these methods and measurements in a dynamic context, deepening our understanding of such stellar systems.

**Keywords:** binary stars · binaries · HIP 45571 · HD 80671

## 1 Introduction

In the vast expanse of the cosmos, many stellar systems are not solitary entities but rather intricate arrangements of multiple star systems. These systems typically involve two gravitationally linked stars, orbiting around a shared center of mass. Understanding the intricacies of binary and multiple stellar systems (BMSSs) is pivotal for delving deeper into stellar astrophysics. By unraveling the characteristics, behaviors, and developmental pathways of BMSSs, we unlock valuable insights that propel our understanding of the universe forward [1].

Scientists estimate various physical and geometrical properties, such as effective temperature, radius, and apparent and absolute magnitudes, to understand binary stars. This is done through computational methods as well as photometric and astrometric observations. Binary stars are important because they provide essential data on mass-luminosity and mass-radius relations (MLR and MRR), respectively, which can only be obtained from binary systems Kallrath & Milone, 2009).

This research delves into a comprehensive investigation of HIP 45571, a stellar entity within the vast celestial expanse. It embarks on a meticulous journey aimed at unraveling its fundamental properties. This pursuit involves precise analysis of its orbital elements and thoroughly examining its physical characteristics. It is a dedicated effort to deeply grasp its inherent attributes within astrophysical phenomena, facilitating a nuanced understanding of the star, its cosmic environment, and its cosmic significance.

The system was recently analyzed by Suhail Masda as a solar metallicity ($Z = 0.019$) binary system. We are reanalyzing the system using the same method, but this time utilizing Gaia DR3 observations and focusing on more details. We considered the mentality of the system as $Z = 0.011$ as reported by [4] and [5] (See Table 1). Furthermore, the Multiple Star Catalog shows that the system is a quadruple one. It consists of a close visual binary AB with a period of 3.445 years (the main focus of this work), and another binary CaCb located 18.834 arcseconds away. The CaCb binary has a separation of 0.557 arcseconds and an orbital period of 86.2266 years. We anticipate the system hierarchy to align with the description provided (see Fig. 1).

In this study, we leveraged four parallaxes obtained from two prominent space missions: the Hipparcos mission, which boasts two published catalogues brimming with relevant system data, and the Gaia mission, known for its three comprehensive data releases. To unravel the orbital solution of the system, we turned to Tokovinin's dynamical method, esteemed for its ability to provide thorough solutions in determining orbital parameters [6].

In estimating the system's physical characteristics, we employed "Al-Wardat's method for analyzing binary and multiple stellar systems (BMSSs)", a sophisticated computational spectrophotometric technique. This method seamlessly builds synthetic spectral energy distributions (SEDs) using the plane-parallel line-blanketed model atmospheres of Kurucz's (ATLAS9) with observational data, thereby unveiling the parameters of the individual components [7–12].

It is worth mentioning that Al-Wardat's method has been used to analyze many binary and multiple stellar systems, for example, see [7, 8, 10, 13–25].

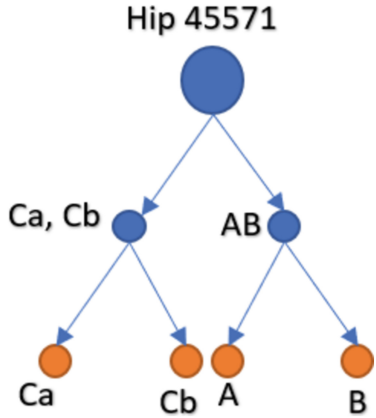

**Fig. 1.** The expected hierarchy of the quadruple system HIP 45571.

## 2  Basic Information of HIP 45571

HIP 45571 (HD 80671) has four different measured parallaxes: the first sourced from the Hipparcos 1997 catalogue ($\pi_{HIP97} = 29.83 \pm 0.60$ mas), followed by the Van Leeuwen catalogue published in 2007 ($\pi_{HIP2007} = 30.64 \pm 0.70$ mas), the third from Gaia data release 2 (DR2) ($\pi_{DR2} = 28.36 \pm 0.28$ mas) [26], and the last from Gaia data release 3 (DR3) ($\pi_{DR3} = 29.46 \pm 0.53$ mas) [5].

The binary system has F5V spectral type according to SIMBAD astronomical database, its effective temperature suggested by [27] is $T_{eff} = 6618$K.

**Table 1.** The basic data of HIP 45571

| Parameter | HIP 45571 | Reference |
| --- | --- | --- |
| $\alpha$ | 09$^h$ 17$^m$ 17$^s$.2424584176 | SIMBAD* |
| $\delta$ | $-68°$ 41$'$ 22$''$.535261160 | SIMBAD* |
| VJ | 5.380 | [30] |
| BT | $5.871 \pm 0.03$ | [31] |
| VT | $5.438 \pm 0.03$ | [31] |
| (B-V) $_{HIP\ old}$ | $0.415 \pm 0.02$ | [31] |
| $\pi_{HIP\ (1997)}$ | $29.830 \pm 0.60$ | [31] |
| $\pi_{HIP\ (2007)}$ | $30.640 \pm 0.70$ | [32] |
| $\pi_{DR2}$ | $28.36 \pm 0.28$ | [26] |
| $\pi_{DR3}$ | $29.46 \pm 0.53$ | [5] |
| $[Fe/H]$ | $-0.25$ | [5] |
| $[Fe/H]$ | $-0.21 \pm 0.05$ | [4] |

The previous orbit parameters of [28] along with modified ones are listed in Table 3. The selection of the stellar system HIP 45571 is based on its expected nature to be a subgiant binary, despite being classified as a main-sequence binary in SIMBAD references. Exploring subgiant stars is crucial due to the scarcity of research in this field. Therefore, we should aim to enhance our understanding and carry out more precise investigations into these celestial systems.

The primary observational data for HIP 45571 is listed in Table 1, along with the corresponding references. These data were sourced from the SIMBAD astronomical database, HIP 97 and HIP 2007 catalogues, Gaia DR2 and DR3. To find the orbital elements, we utilize the relative positional measurements of the two components given in the Fourth Interferometric Catalogue [29].

## 3 Analysis

### 3.1 Orbital Solution

As we mentioned before the system was solved by Tokovinin and his team in 2015 [28], Tokovinin's dynamical method is a computer code that can be used to solve the orbits of binary and triple systems. It can be run under the IDL platform [33]. We used the relative position measurements: angular separations ($\rho$) in (arcsecond), position angles ($\theta$) in (deg), and the relative position measurements of the system's residuals ($\Delta\rho$) in (arcsecond). The new data provided in the fourth interferometric catalogue assisted us in slightly modifying the orbital solution of the system. Figure 2 shows a comparison between the orbit obtained in this work and the previous one. The two solutions are highly consistent.

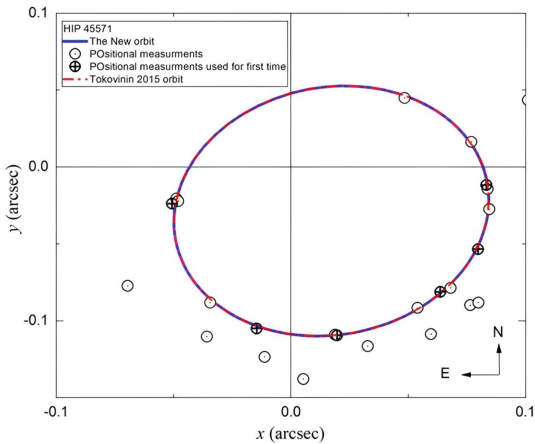

**Fig. 2.** Comparison between the previous orbit [28] and orbit from this work, where we used 5 new relative position measurements (labeled in crossed circles).

### 3.2 Physical Parameters

To estimate the physical parameters of HIP 45571, we used Al-Wardat's method, this method used primary calculation to build synthetic energy distribution (SED) for each component using ATLAS9 [34], the SED produced by ATLAS9 was used to build SED for the entire system using Al-Wardat's combination program. As a double-check, we used synthetic photometry from Al-Wardat's method to compare the synthetic magnitude and the color indices with the observational ones.

The primary calculation includes the difference in visual magnitude ($\Delta m_v = 0.64$) form [29], the apparent visual magnitude for component A and component B ($m_v^A$ and $m_v^B$) of the binary star HIP 45571 calculated using the Eq. (1) for component A:

$$m_v^A = m_v + 2.5 \log(1 + 10^{-0.4 \Delta m_v}) \tag{1}$$

For component B we used the following equation:

$$m_v^B = m_v^A + m_v \tag{2}$$

Next, we calculated the absolute magnitude using the following equation:

$$M_v = m_v + 5 - 5\log(d) - A_v \tag{3}$$

where d is the distance calculated depending on the parallaxes from the past mentioned catalogue and $A_v$ is the interstellar extinction coefficient taken from [35]. To calculate the distances using the parallax we used Eq. (4):

$$d = \frac{1}{\pi} \tag{4}$$

We used Lang Table [36] to obtain the preliminary values of the effective temperatures ($T_{eff}$), the bolometric correction ($BC$), Masses ($m$) of the components and the spectral types. Using the absolute magnitude that we calculated from Eq. 3 and the bolometric correction ($BC$), we can find the preliminary absolute bolometric magnitude ($M_{bol}$) according to the following equation:

$$M_{bol} = M_v + BC \tag{5}$$

we calculated the luminosity ($L$) using the following equation:

$$M_{bol}^* - M_{bol}^\odot = -2.5 \log\left(\frac{L^*}{L^\odot}\right) \tag{6}$$

To calculate Radii ($R$) we used:

$$\frac{R_*}{R_\odot} = \left(\frac{L_*}{L_\odot}\right)^{\frac{1}{2}} \times \left(\frac{T_{eff}^\odot}{T_{eff}^*}\right)^2 \tag{7}$$

The final physical parameter we calculated is the gravitational acceleration (log (g)):

$$\log(g) = \log\left(\frac{M_*}{M_\odot}\right) + \log\left(\frac{R_*}{R_\odot}\right) + 4.43 \tag{8}$$

Given $T_\odot = 5777$ K, $R_\odot = 6.69 \times 10^8$ m, and $M_{bol}^\odot = 4^m.75$ magnitudes, we utilize solar metallicity model atmospheres generated by ATLAS 9 with the preliminary parameters. The total energy flux from a binary star, created by the net luminosities of components $a$ and $b$, situated at a distance $d$ (in parsecs) from Earth, can be calculated as follows [7, 14]:

$$F_\lambda d^2 = H_\lambda^A R_A^2 + H_\lambda^B R_B^2 \tag{9}$$

Rearranging

$$F_\lambda = (R_A^2/d^2)^2 [H_\lambda^A + H_\lambda^B \cdot \left(\frac{R_B}{R_A}\right)^2] \tag{10}$$

$H_\lambda^A$ and $H_\lambda^B$ represent the fluxes from the unit surface of each component, with $F_\lambda$ denoting the total spectral energy density of the system. $R_A$ and $R_b$ stand for the radii of the primary and secondary components in solar units. Synthetic photometry is a common method in various astronomical applications. Al-Wardat's method was employed to synthesize Spectral Energy Distributions, aiding in the calculation of magnitudes and color indices for the binary system.

The method also has a calibration procedure whereby the available observational data was compared with synthetic magnitudes of the complete star system. The accuracy of the results obtained is determined by the level of precision in the observations. Usually, a certain degree of error is acceptable to reach an agreement. The process of fitting colour measurements is used to determine the temperature of a star. By fitting the magnitude differences between the components, it is possible to estimate their relative brightness. Additionally, by fitting visual magnitude measurements between synthetic and observational data, it is possible to measure the overall brightness of the system. These comparative studies significantly contribute to our understanding of binary systems and their stellar components by testing and refining the models produced by this computational method.

The output in Fig. 3 shows the results of the synthetic components (A and B) and overall SEDs of the system Hip 45571 using Al-Wardat's Method and the new parallax ($\pi_{Thiswork} = 28.72 \pm 0.30$ mas).

Stellar metallicity is an important parameter in the analysis of any star, as it describes the abundance of elements heavier than hydrogen and helium in it and is measured by comparing the amount of iron (Fe) to hydrogen (H) in the star on a logarithmic scale relative to the Sun according to the following Equation:

$$\left[\frac{Fe}{H}\right] = \log_{10} \frac{[N_{Fe}/N_H]_{star}}{[N_{Fe}/N_H]_{sun}} \tag{11}$$

where $N_{Fe}$, $N_H$ are the number of iron and hydrogen atoms in a unit of volume. The solar Metallicity was taken as $Z_\odot = 0.0196 \mp 0.0014$ [37].

Gaia DR3 gives the metallicity of Hip 45571 as $\left[\frac{Fe}{H}\right] = -0.25$ (See Table 1). Which means $Z_{Hip45571} = 0.011$. As a part of the method, we tested the positions of the components on the isochrons of different metallicities, and we found it fit the measured metallicity (see Fig. 4). According to Gaia DR3, the metallicity of Hip 45571 is given in Table 1 as $\left[\frac{Fe}{H}\right] = -0.25$, i.e. $Z_{\text{Hip45571}} = 0.011$. As a part of the method's procedures, the positions of the components were tested on the isochrons of different metallicities. It was found that the measured metallicity fit the isochrons of $Z_{\text{Hip45571}} = 0.011$, as shown in Fig. 4.

**Table 2.** Modified orbital parameters of the HIP 45571 along with the previous solution.

| Elements | Units | [28] | Modified elements (This work) |
|---|---|---|---|
| P | [year] | 3.445 | 3.445 ± 0.005 |
| a | [arcsec] | 0.0888 | 0.0888 ± 0.0001 |
| T | – | 2013.376 | 2013.376 ± 0.001 |
| e | – | 0.455 | 0.4550 ± 0.0008 |
| Ω | [deg] | 154.6 | 154.6 ± 0.2 |
| ω | [deg] | 117.2 | 117.20 ± 0.21 |
| i | [deg] | 140.7 | 140.70 ± 0.12 |
| $RMS(\theta)$ | [deg] | – | 0.85 |
| $RMS(\rho)$ | [mas] | – | 0.0009 |

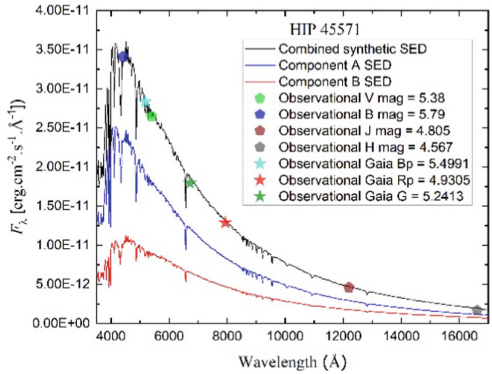

**Fig. 3.** The best fit for the spectral energy distribution (SED) of the Hip 45571 system using observational magnitudes from various sources.

The positions of the system components on the evolutionary tracks and isochrons are strong proof of the reliability of "Al-Wardat's method" as it shows that using the parallax measurements of Hipparcos and Gaia gives the exact solutions and masses. But as part of "Al-Wardat's method," when we combine the orbital solution with estimated masses, we are able to obtain a new dynamical parallax.

Kepler's third law was employed to determine the dynamical mass sum.

$$M_{dyn} = M_A + M_B = \frac{a^3}{\pi^3 P^2} M_\odot \qquad (12)$$

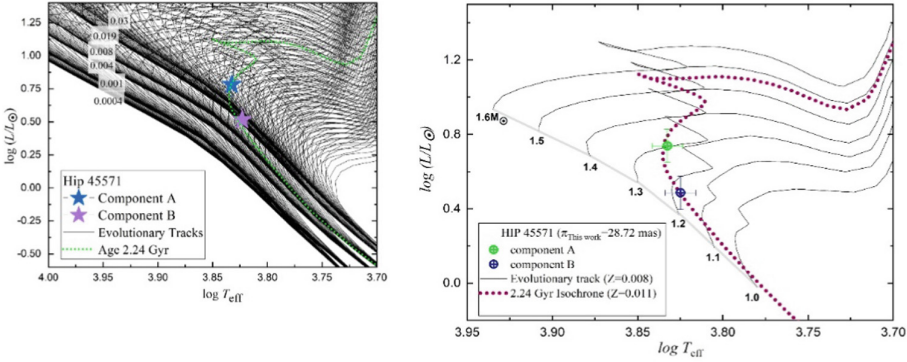

**Fig. 4.** The positions of the system's components on the isochrones and HR diagram and evolutionary tracks of different metallicities as of Girardi et al., 2000. It shows that the system's metallicity Z lies between 0.008 and 0.019.

where $M_A$ represents the mass of the primary component, $M_B$ represents the mass of the secondary component, $a$ denotes the semi-major axis in arcseconds, $\pi$ signifies the parallax in arcseconds and $P$ represents the orbital period in years.

The formula gives the formal error in the total dynamical mass:

$$\frac{\sigma M_{Dyn}}{M_{Dyn}} = \sqrt{9\left(\frac{\sigma_\pi}{\pi}\right)^2 + 9\left(\frac{\sigma_a}{a}\right)^2 + 4\left(\frac{\sigma_P}{P}\right)^2} \tag{13}$$

Now, the important point is finding the system's exact masses and exact parallax.

The method proposed by Al-Wardat provides solutions that are not significantly affected by differences in parallax. Therefore, it is a reliable way to obtain precise masses for individual components. These masses and other estimated parameters are listed in Table 4. By using these masses and modified orbital parameters (Fig. 1), a new parallax value can be obtained as $\pi_{this\ work} = 28.72 \pm 0.30$ mas.

**Table 3.** Comparison between dynamical masses and Al-Wardat's method mass sum of HIP45571.

| Parallax ($\pi$)(mas) | Dynamical mass sum ($M_\odot$) | | Al-Wardat's method mass sum | |
|---|---|---|---|---|
| | [28] | This work | | |
| | | | With Z = 0.008 | With Z = 0.0019 |
| $\pi_{Hip\ 97} = 29.83$ | 2.2228 | $2.22 \pm 0.13$ | $2.48 \pm 0.15$ | $2.78 \pm 0.15$ |
| $\pi_{Hip2007} = 30.64$ | 2.5113 | $2.05 \pm 0.14$ | $2.46 \pm 0.15$ | $2.74 \pm 0.15$ |
| $\pi_{DR2} = 28.36$ | 2.58667 | $2.59 \pm 0.08$ | $2.51 \pm 0.10$ | $2.83 \pm 0.15$ |
| $\pi_{DR3} = 29.46$ | 2.30761 | $2.31 \pm 0.13$ | $2.50 \pm 0.15$ | $2.80 \pm 0.15$ |
| $\pi_{this\ work} = 28.72 \pm 0.30$ | $2.49 \pm 0.07$ | $2.49 \pm 0.07$ | $2.48 \pm 0.10$ | $2.81 \pm 0.15$ |

**Table 4.** The estimated physical parameters for primary and secondary components of HIP 45571.

| Parameter | Units | Comp. A | Comp. B |
|---|---|---|---|
| $T_{eff} \pm \sigma_{T_{eff}}$ | [K] | $6800 \pm 80$ | $6680 \pm 70$ |
| $R \pm \sigma_R$ | $[R_\odot]$ | $1.74 \pm 0.09$ | $1.35 \pm 0.08$ |
| $\log g \pm \sigma_{\log g}$ | $\left[\frac{m}{s^2}\right]$ | $4.19 \pm 0.11$ | $4.50 \pm 0.13$ |
| $L \pm \sigma_L$ | $[L_\odot]$ | $5.78 \pm 0.20$ | $3.23 \pm 0.10$ |
| $M_v \pm \sigma_{M_v}$ | [mag] | $2.81 \pm 0.13$ | $3.43 \pm 0.14$ |
| $M_{bol} \pm \sigma_{M_{bol}}$ | [mag] | $2.80 \pm 0.08$ | $3.47 \pm 0.08$ |
| M | $M_\odot$ | $1.30 \pm 0.15$ | $1.18 \pm 0.13$ |
| Sp. Type | -- | *F*2.5*IV* | *F*3.5*IV* |
| Age | [Gyr] | 2.24 | |

## 4 Conclusion

The purpose of this study was to enhance our understanding of the binary star system called HIP 45571. We achieved this by analyzing its characteristics and physical properties. To refine the orbital parameters, we used Tokovinin's dynamical method in the ORBITX code. This process led to the acquisition of accurate computed orbital elements that are considered to be reliable.

We utilized "Al-Wardat's Method for analyzing BMSSs" to determine physical parameters such as distances, radii, effective temperatures, and gravity. The alignment between synthetic and observational magnitudes and colour indices supported the accuracy of SED predictions, reaffirming the reliability of parallax measurements.

These are the key takeaways that we have concluded on:

- A modified orbit with more accurate orbital elements was achieved by adding five additional relative position measurements (listed in Table 2).
- The physical parameters of the system's components (Table 5) were estimated based on the best fit between synthetic Spectral Energy Distributions (SEDs) and observed photometry.
- We determined that the system comprises *F*2.5*IV* and *F*3.5*IV* solar-type subgiant stars with solar metallicity $Z_* = 0.011$. This conclusion is drawn from analyzing the physical properties of the system's components and their positions relative to evolutionary and age tracks (2.24 Gyr).
- We have determined that the system consists of two subgiant stars, assuring its metallicity $Z_* = 0.011$, which is a little bit smaller than that of the sun. We arrived at this conclusion by analyzing the system's components' physical properties and positions relative to evolutionary tracks and isochrons. We estimate that the system is approximately 2.24 billion years old.
- The analysis resulted in estimating the masses of system components. These estimates, along with the modified orbital elements, produced a new parallax ($\pi_{This work} =$ 28.72 ± 0.30 mas). It is worth noting that this value matches closely with the measurements provided by Gaia DR2. ($\pi_{DR2} = 28.36$ mas), and DR3 ($\pi_{DR3} = 29.46$ mas), Hipparcos 1997 ($\pi_{HIP\,97} = 29.83$ mas), and Hipparcos 2007 ($\pi_{HIP\,2007} = 30.64$ mas).

This comprehensive approach improved our understanding of HIP 45571. It showcased our success in achieving research objectives while emphasizing the importance of precision and reliability when employing these methods and measurements in a dynamic context.

**Acknowledgments.** "Al-Wardat's method for analyzing binary and multiple stellar systems" with its codes, as well as the Fourth Interferometric Catalogue of Binary Stars, SIMBAD database, Sixth Catalog of Orbits of Visual Binary Stars, and ORBITX code, have all been used in his work.